# Superluminal Behaviors of Electromagnetic Near-fields *

WANG Zhi-Yong**, XIONG Cai-Dong

School of Physical Electronics, University of Electronic Science

and Technology of China, Chengdu 610054

Superluminal phenomena have been reported in many experiments of electromagnetic wave propagation, where the superluminal behaviors of evanescent waves are the most interesting ones with the important physical significances. Consider that evanescent waves are related to the near-zone fields of electromagnetic sources, based on the first principles, we study the group velocities of electromagnetic fields in near-field region, and show that they can be superluminal, which can provide a heuristic interpretation for the superluminal properties of evanescent waves.

**PACC:** 4110H, 4210

## 1. Introduction

A series of recent experiments have revealed that electromagnetic wave was able to travel at a group velocity faster than the velocity of light in vacuum, or at a negative group velocity. For example, these phenomena have been observed in dispersive media, [1-4] in electronic circuits, [5] and in evanescent wave cases. [6-13] In fact, over the last decade, the discussion of the tunneling time problem has experienced a new stimulus by the results of analogous experiments with evanescent electromagnetic wave packets, [14-19] and the superluminal effects of evanescent waves have been revealed in photonic tunneling experiments in both the optical domain and the microwave range. [6-13] All the experimental results have shown that the phase time do describe the barrier traversal time. [13, 20]

---

\* Project supported by the Doctoral Program Foundation of Institution of Higher Education of China (Grant No. 20050614022).

\*\* Corresponding author.   E-mail: zywang@uestc.edu.cn

In view of the fact that the evanescent waves actually correspond to the near fields of electromagnetic sources, in this article, we shall show that the group velocities of electromagnetic near-fields can be superluminal, which may provide a heuristic interpretation for the superluminal behaviors reported in many experiments of evanescent wave propagation.

## 2. General expressions of phase velocity and group velocity

There have several different methods to introduce the concepts of group velocity. To give the calculation expression of group velocity form a general point of view, we will choose *the stationary-phase method*. In general, one can expand a field quantity $\Phi(r,t)$ as the superposition of different frequency components with the usual Fourier-transform as its special case ($r = (x, y, z)$):

$$\Phi(r,t) = \frac{1}{(2\pi)^{3/2}} \int_{-\infty}^{+\infty} \Psi(r,k,\omega)\exp[i\Theta(r,t,k,\omega)] \mathrm{d}^3k \qquad (1)$$

where $i = \sqrt{-1}$, $\mathrm{d}^3k = \mathrm{d}k_x \mathrm{d}k_y \mathrm{d}k_z$, $\omega = \omega(k)$ is the frequency, $k = (k_x, k_y, k_z)$ the wave-number vector, $\Psi(r,k,\omega)$ the amplitude of the *k*-th component of $\Phi(r,t)$, and $\Theta(r,t,k,\omega)$ the phase. For example, expanding $\Phi(r,t)$ as the superposition of monochromatic plane waves, one obtains the usual Fourier expansion of $\Phi(r,t)$, for the moment $\Psi(r,k,\omega) = \Psi(k,\omega)$. While, if $\Phi(r,t)$ is expanded as the superposition of monochromatic cylindrical waves (or spherical waves, etc.), the amplitude $\Psi(r,k,\omega)$ would depend on space coordinate $r = (x, y, z)$.

According to the stationary-phase method, the group velocity (say, $v_g$) is taken as the move velocity of the peak of a wave packet, which is valid for both deformed and undeformed wave packets, and in agreement with the phase time theory of tunneling time (note that the phase time has nothing to do with the concept of phase velocity).[15-17] In fact, in quantum mechanics, the classical motion velocity of a particle corresponds to the group velocity of a wave packet, though the wave packet is deformed with time (provided that the particle has nonzero mass). Moreover, according to Ref. [21], the group velocity may be meaningful even for broad band pulses and when the group velocity is superluminal or negative. In the following we denote $a_r = r/r$, $r = |r|$, $k = |k|$ and so on. The superposition of different frequency components gives an extremum at peak location (say, $r_c$) of the wave packet $\Phi(r,t)$, where the peak location $r_c$ is given by $\partial\Theta/\partial k = 0$, and the group velocity is $v_g = a_r \, dr_c/dt$. Let $\Delta \equiv \partial\Theta/\partial k = 0$, one has $d\Delta/dt = (\partial\Delta/\partial r)(dr/dt) + \partial\Delta/\partial t = 0$, and then one has

$$v_g = a_r \, dr_c/dt = -a_r \left(\frac{\partial \Delta}{\partial t}\right) \bigg/ \left(\frac{\partial \Delta}{\partial r}\right) = -a_r \left(\frac{\partial^2 \Theta}{\partial t \partial k}\right) \bigg/ \left(\frac{\partial^2 \Theta}{\partial r \partial k}\right) \quad (2)$$

Eq. (2) is the general expression for calculating the group velocity. As an example, let $\Theta = \omega t - \mathbf{k} \cdot \mathbf{r}$, one has $v_g = a_r \, \partial\omega/\partial k$.

By the way, let us consider that the group velocity along a given direction, say, the z-axis direction. Let $k_z$ denotes the projection of $\mathbf{k}$ in the z-axis direction, it is easy to show that the group velocity along the z-axis direction is

$$v_{gz} = -a_z \left(\frac{\partial^2 \Theta}{\partial t \partial k_z}\right) \bigg/ \left(\frac{\partial^2 \Theta}{\partial z \partial k_z}\right) \quad (3)$$

where $a_z$ is the unit vector along the z-axis direction. Eq. (3) is the most general expression for calculating the group velocity. For example, one can apply Eq. (3) to calculating the group velocity of the electromagnetic waves propagating along a waveguide.

## 3. Group velocities of the near-zone fields of antennas

To gain physics insight, we restrict ourselves to the fields of an electric dipole antenna without loss of generality. For a system of charges and currents varying in time we can make a Fourier analysis of the time dependence and handle each Fourier component separately.[22] We therefore lose no generality by considering the fields of an electric dipole antenna with each frequency component of currents varying sinusoidally in time:

$$\mathbf{J}(\mathbf{x},t) = \mathbf{J}(\mathbf{x})\exp(i\omega t) \quad (4)$$

Note that Eq. (4) just represents a single frequency component of the currents. As usual, the real part of Eq. (4) is to be taken to obtain physical quantities (and so on). For each single frequency component, in the Lorentz gauge, the corresponding vector potential $\mathbf{A}(\mathbf{x},t)$ is

$$\mathbf{A}(\mathbf{x},t) = \frac{\mu}{4\pi} \int \mathbf{J}(\mathbf{x}') \frac{\exp[i(\omega t - \mathbf{k} \cdot \mathbf{r})]}{r} d^3 x' \quad (5)$$

Where $r=|\boldsymbol{r}|=|\boldsymbol{x}-\boldsymbol{x}'|$, $\mu$ is the vacuum permeability. Furthermore, we assume that the electric dipole antenna, as a linear antenna, is oriented along the $z$ axis, extending form $z=-d/2$ to $z=d/2$, and its length $d$ satisfies $d\ll\lambda$, where $\lambda=2\pi c/\omega$ is the wavelength. The magnetic induction is $\boldsymbol{B}=\nabla\times\boldsymbol{A}$ while, outside the source, the electric field is $\boldsymbol{E}=\mathrm{i}\nabla\times\boldsymbol{B}/k$.

On the basis of Ref. [22-23], in spherical coordinate system $(r,\theta,\varphi)$, the non-zero components of the electromagnetic field of the antenna are (for each single frequency component):

$$\begin{cases} H_\varphi = H_0 \dfrac{\sin\theta}{r}(1+\dfrac{1}{ikr})\exp[i(\omega t-kr)] \\ E_r = E_1 \dfrac{\cos\theta}{r^2}(1+\dfrac{1}{ikr})\exp[i(\omega t-kr)] \\ E_\theta = E_2 \dfrac{\sin\theta}{r}(1+\dfrac{1}{ikr}-\dfrac{1}{k^2r^2})\exp[i(\omega t-kr)] \end{cases} \quad (6)$$

Where $\theta$ is the angle between the direction of observation (along $\boldsymbol{r}$) and the polarization direction of the electric dipole moment, $H_0$, $E_1$, $E_2$ are constants. In the near zone of the source, one has $0<kr\ll1$, and from Eq. (6) one has $H_\varphi/E_r\approx0$ and $H_\varphi/E_\theta\approx0$, thus the non-zero components of the electromagnetic field of the antenna can be written as

$$\begin{cases} E_r = E_1 \dfrac{\cos\theta}{r^2}(1+\dfrac{1}{ikr})\exp[i(\omega t-kr)] \equiv |E_r|\exp[i(\Theta+C_1)] \\ E_\theta = E_2 \dfrac{\sin\theta}{ikr^2}(1+\dfrac{1}{ikr})\exp[i(\omega t-kr)] \equiv |E_\theta|\exp[i(\Theta+C_2)] \end{cases} \quad (7)$$

where $C_1$ and $C_2$ are two constants, they are independent of space-time coordinates and have no contribution in calculating the group velocity via Eq. (2). Moreover, $|E_r|$ and $|E_\theta|$ represent the amplitudes of $E_r$ and $E_\theta$, respectively, they also have nothing to do with our topic. To calculate the group velocity via Eq. (2), one must calculate the phase $\Theta$ for the each frequency component, and it is easy to obtain that:

$$\Theta = \omega t - kr + \arctan(-1/kr) \quad (8)$$

where $\arctan(x)$ is the inverse tangent function of $x$. In our case, $\partial\omega/\partial k = c$ corresponds to the velocity of light in vacuum. Using Eqs. (2) and (8), one has:

$$v_g = a_r [\frac{(kr)^4 + 2(kr)^2 + 1}{(kr)^4 + 3(kr)^2}] c = a_r v_g \qquad (9)$$

By Eq. (9) we give the diagrammatic curve of $v_g/c$ (instead of $v_g$) vs $kr$.

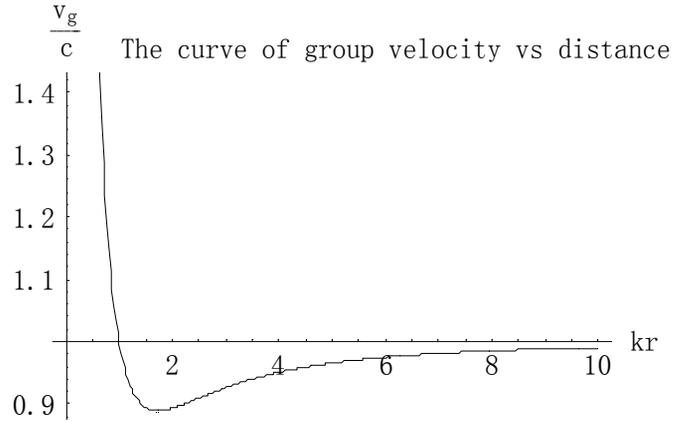

Fig. 1 The curve of group velocity vs distance

Based on Fig. 1, we summarize the main characters of the group velocities of the near fields as follows:

1) In the far zone (for $kr > 1$), the group velocities of the near fields satisfy $8c/9 \approx 0.9c < v_g < c$, and $v_g \to c$ for $kr \to +\infty$. Therefore, in the far zone, the group velocities of the near fields are subluminal;

2) In the near zone (for $0 < kr \leq 1$), contrast to the case 1), the group velocities of the near fields are greater than or equal to the velocity of light in vacuum. Therefore, in the near zone, the group velocities of the near fields are superluminal.

## 4. Relations between near fields and evanescent waves

To associate our theoretical results with those superluminal experiments related to the evanescent waves, we will show that the evanescent waves actually correspond to the near fields of electromagnetic sources. Firstly, there exists a close similarity between the evanescent waves inside an undersize waveguide and the near-zone fields of antennas:

a) Distinct from the travelling waves inside an ordinary waveguide, the evanescent waves inside an undersize waveguide, attenuate exponentially along the direction of propagation, and their average power flows do not exist (but exist as an energy storage), which due to the fact that the impedance is purely capacitive (for the $TM$ mode) or inductive (for the $TE$ mode). As for the energy storage in the undersize waveguide, the electric energy is more than magnetic energy for the $TM$ mode and on the contrary for the $TE$ mode.

b) Similarly, the near-zone fields of an antenna are also sharply attenuated as compared with the far-zone fields of the antenna, and the average power flows of the near-zone fields do not exist because the impedance is purely capacitive (for the electric dipole antenna) or inductive (for the magnetic dipole antenna), it is only an energy storage that exists. As for the energy storage of the near-zone fields, the electric energy is more than magnetic energy for the electric dipole antenna and on the contrary for the magnetic dipole antenna.

In fact, in Ref. [24], the authors (R.P. Feynman et al) have given a detailed analysis for the equivalence between the evanescent waves inside an undersize waveguide and the near fields of a source. Therefore, our conclusion that the near fields have superluminal behaviors is in agreement with the results of the superluminal experiments related to the evanescent waves.

## 5. Conclusions and discussions

On the one hand, the superluminal phenomena of the evanescent waves inside an undersize waveguide have been reported in many experiments; on the other hand, the evanescent waves actually correspond to the near fields of electromagnetic sources, and as demonstrated in this paper, in the near zone of the sources, the group velocities of the near fields are superluminal. Therefore, in this paper we have provided a theoretical interpretation for those superluminal experiments related to the evanescent waves.

In fact, our theoretical results can be understood from the point of view of quantum mechanics. According to quantum mechanics, the uncertainty between the position $r$ and the momentum $p = \hbar k$ ($\hbar$ is the Planck constant), would become remarkable as a measurement is performed within a space interval $r$ satisfying $kr \leq 1$, i.e., within the near zone of sources, such that the superluminal behaviors of the near fields would take place. As quantum-mechanical effects, such superluminal behaviors do not violate the causality principle. However, in this paper we obtain the same conclusions in the level of classical electromagnetic-field theory.

In our next work, we will try to provide a more rigorous analysis for the superluminal behaviors of the near fields, which is based on quantum field theory.

# Acknowledgments

The first author wishes to thank William D. Walker, Prof. Ole Keller as well as Prof. G. Nimtz for valuable discussions.

# References

[1] Bigelow M S, Lepeshkin N N and Boyd R W 2003 *Science* **301** 200.

[2] Kuzmich A, Dogariu A and Wang L J *et al* 2001 *Phys. Rev. Lett*. **86** 3925.

[3] Wang L J, Kuzmich A and Dogariu A 2000 *Nature* **406** 277.

[4] Solli D R, McCormick C F and Ropers C *et al* 2003 *Phys. Rev. Lett.* **91** 143906.

[5] Mitchell M W and Chiao R Y 1998 *Am. J. Phys*. **66** 14.

[6] Steinberg A M, Kwiat P G and Chiao R Y 1993 *Phys. Rev. Lett.* **71** 708.

[7] Enders A and Nimtz G 1992 *J. Phys. I* (France) **2** 1693.

[8] Enders A and Nimtz G 1993 *Phys. Rev. E* **48** 632.

[9] Mugnai D, Ranfagni A and Schulman L S 1997 *Phys. Rev. E* **55** 3593.

[10] Mugnai D, Ranfagni A and Ronchi L 1998 *Phys. Lett. A* **247** 281.

[11] Balcou Ph and Dutriaux L 1997 *Phys. Rev. Lett.* **78** 851.

[12] Alexeev I, Kim K Y and Milchberg H M 2002 *Phys. Rev. Lett*. **88** 073901.

[13] Nimtz G and Heitmann W 1997 *Prog. Quant. Electr.* **21** 81.

[14] Büttiker M and Landauer R 1982 *Phys. Rev. Lett*. **49** 1739.

[15] Hauge E H and Støvneng J A 1989 *Rev. Mod. Phys.* **61** 917.

[16] Olkhovsky V S and Recami E 1992 *Phys. Reports* **214** 339.

[17] Landauer R and Martin Th 1994 *Rev. Mod. Phys*. **66** 217.

[18] Kwiat P G, Chiao R Y and Steinberg A M 1991 *Physica B* **175** 257.

[19] Martin Th and Landauer R 1992 *Phys. Rev. A* **45** 2611.

[20] Th. Hartman Th 1962 *J. Appl. Phys.* **33** 3427.

[21] Peatross J, Glasgow S A and Ware M 2000 *Phys. Rev. Lett*. **84** 2370.

[22] J. D. Jackson J D 1975 *Classical Electrodynamics* (New York: John Wiley & Sons, Inc.) pp 391-395.

[23] Gure B S and Hiziroglu H R 1998 *Electromagnetic Field Theory Fundamentals* (New York: JPWS

Publishing Company) pp 477-479.

[24] Feynman R P, Leighton R B and Sands M 1964 *Feynman Lectures on Physics* Vol. (New York: Addison-Wesley Publishing Company), Chapter 7-5 and Chapter 24-8.